\documentclass[twocolumn]{aastex63}

\usepackage[utf8]{inputenc}
\usepackage{natbib}
\usepackage{graphicx}
\usepackage{amsmath}
\usepackage{booktabs}
\usepackage{longtable}
\usepackage{xspace}
\usepackage{hyperref}
\usepackage[T1]{fontenc}
\usepackage{lipsum}
\usepackage{comment}
\usepackage{xcolor}
\usepackage{multirow}
\usepackage{IEEEtrantools}
\usepackage{amsmath}

\newcommand{\obliquity}{$3.7\pm5.0^{\circ}$}
\newcommand{\vsini}{$1.6\pm0.3$}
\newcommand{\coplanar}{$12^{\circ}$}

\begin{document}

\title{The HD 191939 Exoplanet System is Well-Aligned and Flat}

\author[0000-0001-8342-7736]{Jack Lubin}
\affiliation{Department of Physics \& Astronomy, University of California Los Angeles, Los Angeles, CA 90095, USA}

\author[0000-0003-0967-2893]{Erik A. Petigura}
\affiliation{Department of Physics \& Astronomy, University of California Los Angeles, Los Angeles, CA 90095, USA}

\author[0000-0002-4290-6826]{Judah Van Zandt}
\affiliation{Department of Physics \& Astronomy, University of California Los Angeles, Los Angeles, CA 90095, USA}

\author[0000-0001-7708-2364]{Corey Beard}
\altaffiliation{NASA FINESST Fellow}
\affiliation{Department of Physics \& Astronomy, University of California Irvine, Irvine, CA 92697, USA}

\author[0000-0002-8958-0683]{Fei Dai}
\affiliation{Institute for Astronomy, University of Hawai`i, 2680 Woodlawn Drive, Honolulu, HI 96822, USA}
\affiliation{Division of Geological and Planetary Sciences,
1200 E California Blvd, Pasadena, CA, 91125, USA}
\affiliation{Department of Astronomy, California Institute of Technology, Pasadena, CA 91125, USA}

\author[0000-0003-1312-9391]{Samuel Halverson}
\affiliation{Jet Propulsion Laboratory, California Institute of Technology, 4800 Oak Grove Drive, Pasadena, California 91109}

\author[0000-0002-5034-9476]{Rae Holcomb}
\affiliation{Department of Physics \& Astronomy, University of California Irvine, Irvine, CA 92697, USA}

\author[0000-0001-8638-0320]{Andrew W.\ Howard}
\affil{Department of Astronomy, California Institute of Technology, Pasadena, CA 91125, USA}

\author[0000-0002-0531-1073]{Howard Isaacson}
\affiliation{Department of Astronomy, 501 Campbell Hall, University of California, Berkeley, CA 94720, USA}
\affiliation{Centre for Astrophysics, University of Southern Queensland, Toowoomba, QLD, Australia}

\author[0000-0002-4927-9925]{Jacob Luhn}
\affiliation{Department of Physics \& Astronomy, University of California Irvine, Irvine, CA 92697, USA}

\author[0000-0003-0149-9678]{Paul Robertson}
\affiliation{Department of Physics \& Astronomy, University of California Irvine, Irvine, CA 92697, USA}

\author[0000-0003-3856-3143]{Ryan A. Rubenzahl}
\altaffiliation{NSF Graduate Research Fellow}
\affil{Department of Astronomy, California Institute of Technology, Pasadena, CA 91125, USA}

\author[0000-0001-7409-5688]{Gu{\dh}mundur Stef{\'a}nsson}
\affil{Anton Pannekoek Institute for Astronomy, University of Amsterdam, Science Park 904, 1098 XH Amsterdam, The Netherlands}

\author[0000-0002-4265-047X]{Joshua N.\ Winn}
\affiliation{Department of Astrophysical Sciences, Princeton University, Princeton, NJ 08544, USA}


\author[0009-0008-9808-0411]{Max Brodheim}
\affiliation{W. M. Keck Observatory, Waimea, HI 96743, USA}

\author[0009-0000-3624-1330]{William Deich}
\affiliation{UC Observatories, University of California, Santa Cruz, CA 95064, USA}

\author[0000-0002-7648-9119]{Grant M.\ Hill}
\affiliation{W. M. Keck Observatory, Waimea, HI 96743, USA}

\author[0009-0004-4454-6053]{Steven R. Gibson}
\affiliation{Caltech Optical Observatories, California Institute of Technology, Pasadena, CA 91125, USA}

\author[0000-0002-6153-3076]{Bradford Holden}
\affiliation{UC Observatories, University of California, Santa Cruz, CA 95064, USA}

\author[0000-0002-5812-3236]{Aaron Householder}
\affiliation{Department of Earth, Atmospheric and Planetary Sciences, Massachusetts Institute of Technology, Cambridge, MA 02139, USA}
\affil{Kavli Institute for Astrophysics and Space Research, Massachusetts Institute of Technology, Cambridge, MA 02139, USA}

\author[0000-0003-2451-5482]{Russ R. Laher}
\affiliation{NASA Exoplanet Science Institute/Caltech-IPAC, 1200 E California Blvd, Pasadena, CA 91125, USA}

\author[0009-0004-0592-1850]{Kyle Lanclos}
\affiliation{W. M. Keck Observatory, Waimea, HI 96743, USA}

\author[0009-0008-4293-0341]{Joel Payne}
\affiliation{W. M. Keck Observatory, Waimea, HI 96743, USA}

\author[0000-0001-8127-5775]{Arpita Roy}
\affiliation{Astrophysics \& Space Institute, Schmidt Sciences, New York, NY 10011, USA}

\author[0000-0001-7062-9726]{Roger Smith}
\affiliation{Caltech Optical Observatories, California Institute of Technology, Pasadena, CA 91125, USA}

\author[0000-0003-3133-6837]{Abby P. Shaum}
\affiliation{Department of Astronomy, California Institute of Technology, Pasadena, CA 91125, USA}

\author[0000-0002-4046-987X]{Christian Schwab}
\affiliation{School of Mathematical and Physical Sciences, Macquarie University, Balaclava Road, North Ryde, NSW 2109, Australia}

\author[0000-0002-6092-8295]{Josh Walawender}
\affiliation{W. M. Keck Observatory, Waimea, HI 96743, USA}

\begin{abstract}
    We report the sky-projected spin-orbit angle $\lambda$ for HD 191939 b, the innermost planet in a 6 planet system, using Keck/KPF to detect the Rossiter-McLaughlin (RM) effect. Planet b is a sub-Neptune with radius 3.4 $\pm$ 0.8 R$_{\oplus}$ and mass 10.0 $\pm$ 0.7 M$_{\oplus}$ with an RM amplitude $<$1 m\,s$^{-1}$. We find the planet is consistent with a well-aligned orbit, measuring $\lambda= \, $\obliquity. Additionally, we place new constraints on the mass and period of the distant super-Jupiter, planet f, finding it to be 2.88 $\pm$ 0.26 $M_J$ on a 2898 $\pm$ 152 day orbit. With these new orbital parameters, we perform a dynamical analysis of the system and constrain the mutual inclination of the non-transiting planet e to be smaller than \coplanar\ relative to the plane shared by the inner three transiting planets. Additionally, the further planet f is inclined off this shared plane, the greater the amplitude of precession for the entire inner system, making it increasingly unlikely to measure an aligned orbit for planet b. Through this analysis, we show that this system's wide variety of planets are all well-aligned with the star and nearly co-planar, suggesting that the system formed dynamically cold and flat out of a well-aligned proto-planetary disk, similar to our own solar system.
\end{abstract}
\keywords{exoplanet, obliquity, Rossiter-McLaughlin}

\section{Introduction}
\label{Intro}

\par While there are just over 200 spin-orbit angles measured for exoplanets, more than 80\% of them are for Jovians on short orbits (R $>$ 0.5 $R_J$, P $<$ 20d) \citep{Southworth2011, Albrecht2022}. The bias towards short-period giants comes in large part from instrumental and scheduling limitations. The Rossiter-McLaughlin (RM) anomaly \citep{Rossiter1924, McLaughlin1924} amplitude scales directly with planet cross-sectional area, and the precision limits on the prior generation of spectrographs have prevented the measurement of the small anomalies from small planets. Similarly, short orbital period planets that transit frequently and with short durations are easiest to schedule due to the necessity of observing within a single night during transit. Regardless, the results from these measurements are scientifically worthwhile, providing insights into Jovian formation and migration.

\begin{figure*}[t!]
    \centering\includegraphics[width=0.95\textwidth]{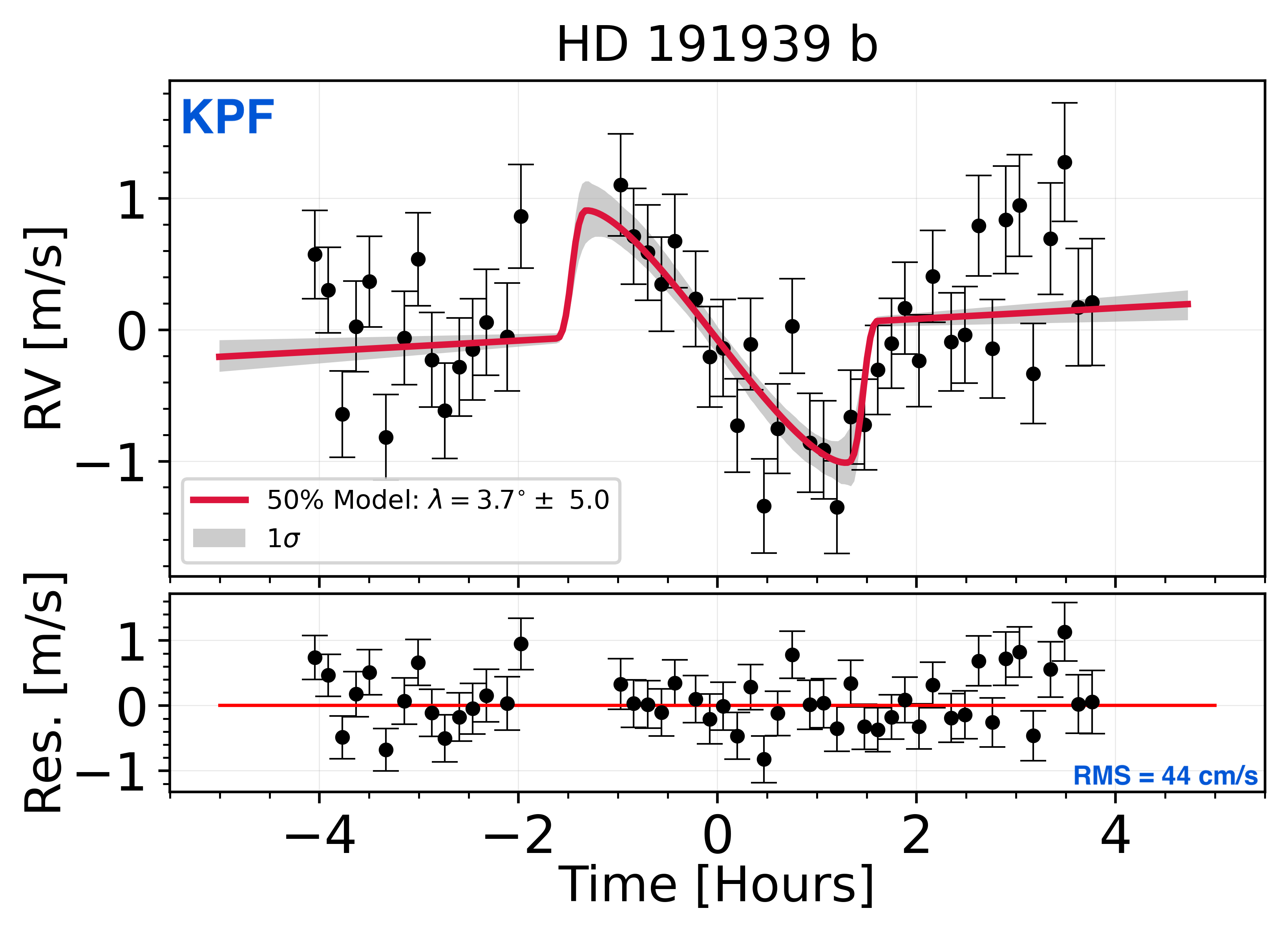}
    \caption{Our KPF RV time series from the night of UT 2023 July 28 (black) and the model of the RM anomaly computed from the median parameter values (red) with residuals in the bottom panel. The RMS of the residuals is 44 cm\,s$^{-1}$}
    \label{fig:rmplot}
\end{figure*}

\par However, due to this observational bias we have severely under-sampled the spin-orbit angles of the very planets that are the most common across the galaxy at short periods: those under 4 $R_{\oplus}$. To date, there have been 16 such planets with measured obliquities from Rossiter-McLaughlin observations, of which 10 are found to be consistent with an aligned orbit: K2-25 b \citep{Stefansson2020, Gaidos2020}, HD 63433 b \citep{Mann2020}, TRAPPIST-1 b/e/f \citep{Hirano2020}, TOI-942 b \citep{Wirth2021}, HD 3167 b \citep{Bourrier2021}, TOI-2076 b \citep{Frazier2023}, 55 Cnc e \citep{Zhao2023}, and HD 110067 b \citep{Zak2024}. Furthermore, the additional 5 small planets with obliquity measurements are inconsistent with alignment: Kepler-408 \citep{Kamiaka2019}, $\pi$ Men c \citep{Kunovac2021}, HD 3167 c \citep{Bourrier2021}, K2-290 b \citep{Hjorth2021}, GJ 436 b \citep{Bourrier2022}, and GJ 3470 b \citep{Stefansson2022}. The diversity in measured obliquities reflects the diversity of small planets and the system architectures they reside in. It is crucial that we make more measurements of these kinds of planets in order to build a sample size large enough to uncover the trends which may hint at the formation and evolution of these planets.

\par With new extreme precision radial velocity (EPRV) instruments, we now have instruments capable of measuring the spin-orbit angles of these sub-Neptune planets. This opens up the possibility of investigating the formation, migration and dynamical histories of new kinds of systems. For example, sub-Neptunes are known to exist in high multiplicity, peas-in-a-pod architectures \citep{Weiss2018}, in contrast to the Hot Jupiters which are most often lonely \citep{Wu2023}. Furthermore, there seems to exist a break in planet occurrence \citep{Fulton2017} and stellar parameters \citep{Buchhave2012} among others at 4$R_{\oplus}$, therefore it is natural to expect the obliquity distribution to change at this planet size at well. By measuring the spin-orbit angles of sub-Neptunes, we are fundamentally probing different evolution and dynamical pathways compared to Hot Jupiters.

\par The HD 191939 system is one such system with high multiplicity of sub-Neptunes that is amenable to detailed characterization. The system consists of 6 total planets. The inner three are transiting sub-Neptunes (9d, 28d, and 38d orbital periods for b, c, and d, respectively) that were first identified and validated with TESS photometry \citep{BadenasAgusti2020}. Then \citet{Lubin2022} followed up the system with radial velocity (RV) measurements from Keck/HIRES and the Automated Planet Finder (APF) and measured the masses of the transiting sub-Neptunes. They further announced the discovery of a non-transiting 101d Warm Saturn (planet e) and performed a dynamical analysis of the system. \citet{Lubin2022} also identified a strong trend and curvature in the RV residuals implying the existence of a 5th body in the system. Through a joint RV and astrometry analysis, further detailed in \citet{VanZandt2024}, \citet{Lubin2022} constrained the mass to be between 2 and 11 $M_J$, the orbital period to between 1200 and 7200 days, and designated it planet f. Finally, \citet{OrellMiquel2023} added additional CARMENES and HARPS-N RVs, which they used to further constrain the known planet masses as well as announce the discovery of an additional sub-Neptune (planet g) on a 280 day orbit.

\par In this work, we add to the small but growing census of sub-Neptunes with measured spin-orbit angles through observations of HD 191939 b with Keck/KPF. We also expand upon previous work to investigate the wider dynamics of this unique system. This work further highlights the benefits of characterizing a system in detail which is highly amenable to follow-up from a variety of techniques, and motivates the need for similar detailed analyses on more such systems.

\par This work is organized as follows. In \S\ref{Observations}, we describe the observational data from Keck/KPF used to measure the RM anomaly. Next, in \S\ref{Spin-Orbit} we analyze the in-transit RVs to measure the spin-orbit angle for planet b. In \S\ref{newmodel}, we re-model the entire six planet system with all archival RVs and two new HIRES observations that help constrain planet f's orbit and mass. Next in \S\ref{Dynamics} we carry out N-body simulations to investigate the dynamics of this constrained system and discuss the implications of this study in \S\ref{Discussion}. Finally we conclude in \S\ref{Conclusion}.

\begin{table}[t]
  \centering
  \caption{Compiled Stellar Parameters}
  \label{stellparams}
  \begin{tabular}{llll}
    \hline
Parameter & Unit & Value & Reference  \\
    \hline
    \hline
Aliases & - & TOI-1339, HIP 99175 &  \\
RA & h:m:s & 20:08:06.15 & G \\
Dec & d:m:s & +66:51:01.08 & G \\
V mag & - & 8.97 & BA20 \\
Mass & M$_{\odot}$ & 0.81 $\pm$ 0.04 & L22 \\
Radius & R$_{\odot}$ & 0.94 $\pm$ 0.02 & L22 \\
Luminosity & L$_{\odot}$ & 0.65 $\pm$ 0.02 & L22 \\
Teff & K & 5348 $\pm$ 100 & L22 \\
log \textit{g} & cm s$^{-2}$ & 4.3 $\pm$ 0.1 & L22 \\
Fe/H & dex & -0.15 $\pm$ 0.06 & L22 \\
$v\sin i$ & km s$^{-1}$ & \vsini & This work \\
Age & Gyr & $>8.7$ & L22 \\
\hline
\hline
\end{tabular}
\footnotesize{G - \cite{Gaia2018}, BA - \citet{BadenasAgusti2020}, L22 - \citet{Lubin2022}}
\end{table}

\section{Observations With KPF}
\label{Observations}

\par We observed HD 191939 with the Keck Planet Finder (KPF) instrument \citep{Gibson2018} at Keck Observatory on UT 2023 July 28. KPF is a fiber-fed EPRV spectrograph that is ultra-stable mechanically and thermally. The main spectrometer has a spectral bandpass of 445--870 nm at a resolving power of $R\sim98,000$. A dedicated Calcium H\&K spectrograph covers the 382--402~nm range at $R\sim17,000$.

\par We used an exposure time of 444 seconds, chosen to average over the expected pressure-mode timescale \citep{Chaplin2019},  see Table \ref{stellparams} for a compiled list of stellar parameters. We obtained a total of 50 spectra over an $\sim$8 hour span of observations from UT 06:58 to 14:55. These observations began at airmass = 1.76, maxed at airmass =1.47 shortly after ingress when the star crossed the local meridian, and observations ended at sunrise at airmass = 2.76. Of these, 17 spectra were obtained within the 3 hour transit window between 09:48 UT and 12:43 UT, all taken at airmass $<$ 1.55. Before ingress, we obtained 16 spectra and after egress we obtained another 17 spectra. Using the updated ephemeris information \citep{OrellMiquel2023} and the Transit and Ephemeris Service tool from the NASA Exoplanet Archive, we predicted the transit midpoint to be at UT2023-07-28:11:15 = BJD 2460153.96875, with a propagated uncertainty of 53 seconds.

\par Conditions were highly favorable on the night of the transit event. However, after electing to take a focus and a calibration exposure just before expected ingress, when we returned to the target star the tip/tilt corrector which holds the star on the science fiber temporarily failed. Our skilled observers caught this error and our telescope operator was able to reset the system after $\sim$40 min of real-time debugging. However, after careful inspection in the hope of recovering these spectra, we were forced to throw them all away due to low counts when the star had drifted off the fiber. These 6 spectra are not included in the accounting above. Unfortunately these 6 observations, spanning UT 09:26 to 10:07, include the expected ingress and represent the gap in observations in Figure \ref{fig:rmplot}.

\par The 50 spectra used in our analysis were reduced in the standard KPF Data Reduction Pipeline\footnote{\url{https://github.com/Keck-DataReductionPipelines/KPF-Pipeline/}}, which makes use of the Cross-Correlation Function (CCF) technique. See \citet{Rubenzahl2023} for more details on the reduction pipeline.

\begin{figure}[t!]
\centering\includegraphics[width=0.48\textwidth]{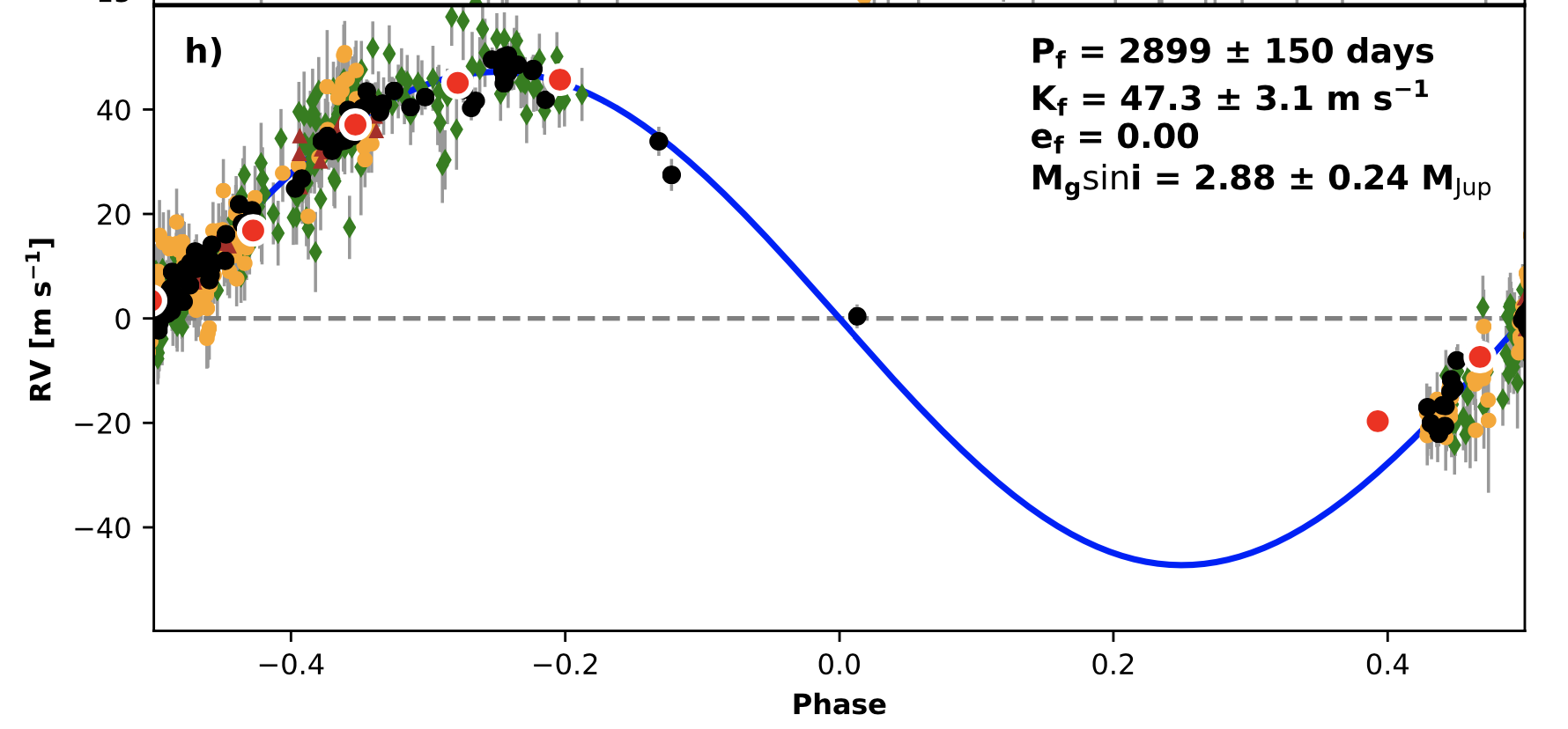}
    \caption{Phase-folded RV curve for the distant super-Jovian, planet f, from the 6 planet model (other planets have been subtracted). Full priors and posteriors can be found in Table \ref{tab:priorsposteriors}}
    \label{fig:planetf}
\end{figure}

\section{Spin-Orbit Angle Measurement}
\label{Spin-Orbit}

\par We elected to model the the RM anomaly with the \texttt{rmfit} \citep{Stefansson2022} package which employs the framework from \citet{Hirano2011}. We fixed the transit parameters of period, the transit midpoint, inclination, $R_p$/$R_*$, a/$R_*$, and eccentricity equal to the median posterior values of the updated lightcurve model from \citet{OrellMiquel2023}. Next we placed Gaussian priors on the the limb-darkening parameters, based on calculations using the Exoplanet Characterization Toolkit \citep{Bourque2021} over the bandpass of KPF and median values for the stellar parameters from \citet{OrellMiquel2023}. Next, we set a uniform prior for the sky-projected spin-orbit angle $\lambda$, \, $\mathcal{U}$(-180, 180) as well as a uniform prior on the stellar $v\sin$i \, $\mathcal{U}$(0, 5), as both \citet{Lubin2022} and \citet{BadenasAgusti2020} found the $v\sin i$ is less than 2 km\,s$^{-1}$ from analysis of the HIRES and NRES spectra, respectively.

\par We applied a uniform prior on the RV slope over the observation window. Over the 8 hour set of observations there will be a Doppler shift of the star due to the planet's true motion through its orbit during the time of the observations. Because this is a multi-planet system with many true orbital motions at play, this term is interpreted as a cumulative trend to the data. This trend also encompasses any instrumental effects and/or stellar variability effects that may cause a slope in the RVs during the observations. Through all this, we set the prior as $\mathcal{U}$(-10, 10) so as to allow for both positive and negative slopes which may be the result of the combination of all these effects.

\par We ran Markov Chain Monte Carlo (MCMC) sampling to explore the full parameter space and estimate error bars. After identifying a global maximum-likelihood with the \texttt{PyDE} differential evolution optimizer \citep{pyde}, the MCMC took 30,000 steps using \texttt{emcee} \citep{emcee}. The Gelman-Rubin \citep{Ford06} statistic for each free parameter was less than 1.01 and so we considered the sampling to be converged. The posterior values for each parameter can be found in Table \ref{allparams} and the resulting corner plot is in the appendix. Most notably, we found the sky-projected spin-orbit angle to be \obliquity \, degrees and the stellar $v\sin i$ to be \vsini \, km\,s$^{-1}$.

\begin{deluxetable*}{llll}
\label{allparams}
\tablecaption{Priors and posteriors for spin-orbit model of planet b and full RV model of the 6 planet system 
\label{tab:priorsposteriors}}
\tablehead{ & \colhead{Description (units)} &\colhead{Priors} &\colhead{Posterior}}
\startdata
\multicolumn{4}{l}{\textbf{RM Parameters:}}\\
$\lambda$ & Sky-projected spin-orbit angle (deg)      & $\mathcal U(-180,180)$   &   \obliquity  \\
$v \sin{i_*}$ & Host star projected rotational velocity (km/s)      & $\mathcal U(0,5)$   &  \vsini \\
$K$& Radial velocity slope (m/s)  & $\mathcal U(-10,10)$   &   $-1.4\pm0.8$  \\
$\beta$ & Intrinsic stellar line width (km/s) & $\mathcal G(3.0,2.0)$ &  $3.3\pm1.9$ \\
$q_{1,\rm KPF}$ & Linear limb-darkening coefficient for KPF$^{a}$&  $\mathcal G(0.55,0.5)$   &  $0.34\pm0.40$  \\
$q_{2,\rm KPF}$ & Quadratic limb-darkening coefficient for KPF$^{a}$&  $\mathcal G(0.22,0.5)$   &  $0.08\pm0.42$  \\
\\
\multicolumn{2}{l}{\textbf{Planetary Parameters for RM model:}} & \multicolumn{2}{l}{}\\
$P_b$& Orbital period (days) & 8.8803256  &   Fixed \\
$T_{0;b}$& Transit epoch (BJD) & 2460153.968750 &  Fixed \\
$R_b / R_\star$ & Planet-to-star radius ratio & 0.03319 &  Fixed \\
$a_b / R_\star$ & Orbital separation over stellar radius & 18.36 & Fixed \\
$i_b$ & Orbital inclination (degrees) & 88.10 & Fixed \\
\\
\multicolumn{2}{l}{\textbf{Planetary Parameters for system model$^{b}$:}} & \multicolumn{2}{l}{}\\
$P_b$ & Orbital period (days) & 8.88032  & Fixed  \\
$T_{0;b}$ & Transit epoch (BJD) & 2458715.356133 & Fixed \\
$K_{b}$ & Amplitude (m/s) & $\mathcal U(0,1000)$  &  3.44 $\pm$ 0.24 \\
$M_{b}$ & M$_{\oplus}$ & Derived  &  9.66 $\pm$ 0.7 \\
SM Axis$_b$ & AU & Derived  &  0.078 $\pm$ 0.0009 \\
$P_c$ & Orbital period (days) & 28.5805  & Fixed  \\
$T_{0;c}$ & Transit epoch (BJD) & 2458726.053366 & Fixed \\
$K_{c}$ & Amplitude (m/s)& $\mathcal U(0,1000)$  &  1.82 $\pm$ 0.25 \\
$M_{c}$ & M$_{\oplus}$ & Derived  &  7.55 $\pm$ 1.05 \\
SM Axis$_c$ & AU & Derived  &  0.17 $\pm$ 0.002 \\
$P_d$ & Orbital period (days) & 38.3524  & Fixed  \\
$T_{0;d}$ & Transit epoch (BJD) & 2458743.551787 & Fixed \\
$K_{d}$ & Amplitude (m/s) & $\mathcal U(0,1000)$  &  0.61 $\pm$ 0.24 \\
$M_{d}$ & M$_{\oplus}$ & Derived  &  2.80 $\pm$ 1.11 \\
SM Axis$_d$ & AU & Derived  &  0.21 $\pm$ 0.002 \\
$P_e$ & Orbital period (days) & $\mathcal G(101.5,2.0)$  &  101.7 $\pm$ 0.08 \\
$T_{0;e}$ & Time of Conjunction (BJD) & $\mathcal U(2459000,2459100)$ & 2459044.06 $\pm$ 0.26 \\
$K_{e}$ & Amplitude (m/s) & $\mathcal U(0,1000)$  &  18.0 $\pm$ 0.25 \\
$M_{e}$ & M$_{\oplus}$ & Derived  & 114.1 $\pm$ 3.1 \\
SM Axis$_e$ & AU & Derived  &  0.40 $\pm$ 0.004 \\
$P_g$ & Orbital period (days) &  $\mathcal G(284,10)$  &  288.6 $\pm$ 7.1 \\
$T_{0;g}$ & Time of Conjunction (BJD) & $\mathcal U(2459372,2459398)$ & 2459392.3 $\pm$ 7.1 \\
$K_{g}$ & Amplitude (m/s) & $\mathcal U(0,1000)$  &  0.77 $\pm$ 0.28  \\
$M_{g}$ & M$_{\oplus}$ & Derived  &  6.88 $\pm$ 2.50 \\
SM Axis$_g$ & AU & Derived  &  0.80 $\pm$ 0.016 \\
$P_f$ & Orbital period (days) & $\mathcal U(1000,10000)$  &  2898 $\pm$ 152 \\
$T_{0;f}$ & Time of Conjunction (BJD) & $\mathcal U(2458400,2462400)$ & 2460449 $\pm$ 43 \\
$K_{f}$ & Amplitude (m/s) & $\mathcal U(0,1000)$  &  47.3 $\pm$ 3.3 \\
$M_{f}$ & M$_{Jup}$ & Derived  &  2.88 $\pm$ 0.26 \\
SM Axis$_f$ & AU & Derived  &  3.71 $\pm$ 0.13 \\
\enddata
\tablenotetext{a}{Initial estimates for limb-darkening parameters come from \citet{exoctk}}
\tablenotetext{b}{For system model, all planets' eccentricities and arguments of periastron were fixed to zero.}
\end{deluxetable*}

\section{Re-Modeling the Full System}
\label{newmodel}

\par Since the publication of \citet{Lubin2022}, who reported 73 HIRES RVs and 104 APF RVs, many more RVs have been taken with multiple instruments. First, \citet{OrellMiquel2023} add 42 HARPS-N RVs as well as 138 CARMENES RVs. Recently, \citet{Polanski2024} published an additional 37 HIRES RVs and an additional 150 APF RVs as part of the full data release of the TESS-Keck Survey (TKS) \citep{Chontos2022}. Additionally, there are two HIRES RVs taken in May 2023 and one in June 2024 that were unpublished until this work, see Table \ref{tab:hiresRVs}.

\par With so many more additional data, now a total of 544 RVs spanning a total baseline of 1301 days, we used \texttt{radvel} to re-fit the complete data set with the primary goal of constraining the period and mass of the distant super-Jovian, planet f. For the three transiting planets (b, c, and d), we fixed the period and time of conjunction to their median posterior values in \citet{OrellMiquel2023}. For the three non-transiting planets (e, f, and g), we set a Gaussian prior on both period and time of conjunction using the posterior values from \citet{OrellMiquel2023}. For all planets, we fixed the eccentricity to zero as \citet{Lubin2022} finds the circular models to be preferred and \citet{OrellMiquel2023} find small eccentricities, $<$0.03, which is negligible for our goal of modeling planet f's orbital period and mass. We further enforced a positive value on the K-amplitude. We do not allow for any {\it ad hoc} trend or curvature, so as to force the model to account for any such trends as the effects of planet f. We further allowed uniform priors on a jitter term for each instrument, $\mathcal{U}$(0,10) and allowed for the instrumental offsets to be determined by the model.

\par The resulting model found all parameters for planets b, c, d, e, and g to be fully consistent with those found by the preferred model from \citet{OrellMiquel2023}. Additionally, we found that the distant super-Jovian, planet f, has an orbital period of 2898 $\pm$ 142 days, corresponding to a semi-major axis of 3.71 $\pm$ 0.13 AU, with a time of conjunction of BJD $2460449 \pm 43$. Furthermore, we measured its K-amplitude to be 47.3 $\pm$ 3.3 m\,s$^{-1}$, which corresponds to a $M\sin i$ of 2.88 $\pm$ 0.26 $M_J$, confirming its planetary nature (see Figure \ref{fig:planetf}).

\section{Dynamical Modeling}
\label{Dynamics}

\par \citet{Lubin2022} constrained the mutual inclination between the orbital plane of the non-transiting planet e with the roughly shared orbital plane of the inner three transiting sub-Neptunes with Laplace-Lagrange secular perturbation theory \citep{LaplaceOriginal} through investigating how different architectures change how often or not Earth observers see all three planets inner as transiting. With the mass and period of planet f constrained, we can extend this analysis to include planet f and further marginalize the 3-transit probability over all lines of sight. Because planet g is very low mass compared to planets e and f, we choose to ignore it in this analysis as its gravitational effect on the changing inclinations of planets b, c, and d is negligible; and since it is non-transiting it adds no information to constraining the shared orbital plane of the system.

\par We run a suite of N-body simulations using \texttt{Rebound} \citep{rebound, reboundias15}. We initialize a system with planets b, c, and d at their respective semi-major axes, median masses as measured in \citet{OrellMiquel2023}, zero eccentricity, inclination of 90$^{\circ}$, and longitude of ascending node of 0$^{\circ}$. We add planets e and f similarly, except in each simulation, e and f are initialized at a different inclination. We run simulations for all combinations of inclinations of e and f ranging from 90$^{\circ}$ down to and including 70$^{\circ}$, 440 simulations in total. Each simulation is run for one eigen-frequency of the slowest oscillation frequency, for planet e of 85,000 years \citep{Lubin2022}, in steps of 10 years. At each timestamp in each simulation, all 5 planets' inclinations and longitudes of ascending node are recorded.

\par Next, at each timestamp in a single simulation we generated analytic boundaries to the transit shadows of all three inner planets on a 3D unit sphere (see \citet{RagozzineHolman} Figure 1). We then performed a Monte Carlo estimation of the overlapping area of all three transit shadows. We randomly generated coordinates on the surface of the unit sphere by drawing a longitude value from a uniform distribution, $\mathcal{U}(-180, 180)$ and drawing a latitude value from a uniform distribution in $\cos \theta$. At each timestamp in a simulation, 10,000 points are generated and the number of points that fall in all three transit shadows are recorded. Then an average number of points in all three transit shadows is taken for all timestamps within one simulation and recorded as the average probability for a randomly-located observer to see all three planets as transiting planets for that simulation. In effect, this process marginalizes the inherent probability of seeing all three planets being observed as transiting over all lines of sight, or the universal three transit probability. This is repeated for every simulation and results are shown in Figure \ref{fig:marginalizedTransitProb}.

\begin{deluxetable*}{lllll}
\label{allparams}
\tablecaption{Sample$^{a}$ of Observations \label{tab:hiresRVs}}
\tablehead{\colhead{BJD} &\colhead{RV (m/s)} &\colhead{e$_{RV}$ (m/s)} &\colhead{Instrument} &\colhead{Source$^{b}$}}
\startdata
2458794.29057 & -18.23 & 3.23 & CARMENES & A \\
2458795.83217 & -31.87 & 1.22 & HIRES & B \\
2458834.64674 & -31.67 & 3.64 & APF & B \\ 
2459012.68918 & -9225.47 & 0.68 & HARPS-N & A \\
2459803.08254 & 25.52 & 1.82 & HIRES & C \\
2459803.93114 & 22.19 & 2.38 & APF & C \\
2460068.11741 & 3.92  & 1.90 & HIRES & This work$^{c}$ \\
2460095.04327 & -10.19 & 2.27 & HIRES & This work$^{c}$ \\
2460487.88545 & -31.43 & 0.99 & HIRES & This work$^{c}$ \\
\enddata
\tablenotetext{a}{A full machine-readable version is available online.}
\tablenotetext{b}{Sources: A = \citet{OrellMiquel2023}, B = \citet{Lubin2022}, C = \citet{Polanski2024}}
\tablenotetext{c}{While this is a subset of the full data set, these three are the only RVs used in this study that have not been previously published.}
\end{deluxetable*}

\section{Discussion}
\label{Discussion}

\par Our measurement of HD 191939 b's spin-orbit angle, as well as our updated mass and semi-major axis of planet f, allow for three insights into the system as a whole.

\par
\begin{figure}[t!]
\centering\includegraphics[width=0.45\textwidth]{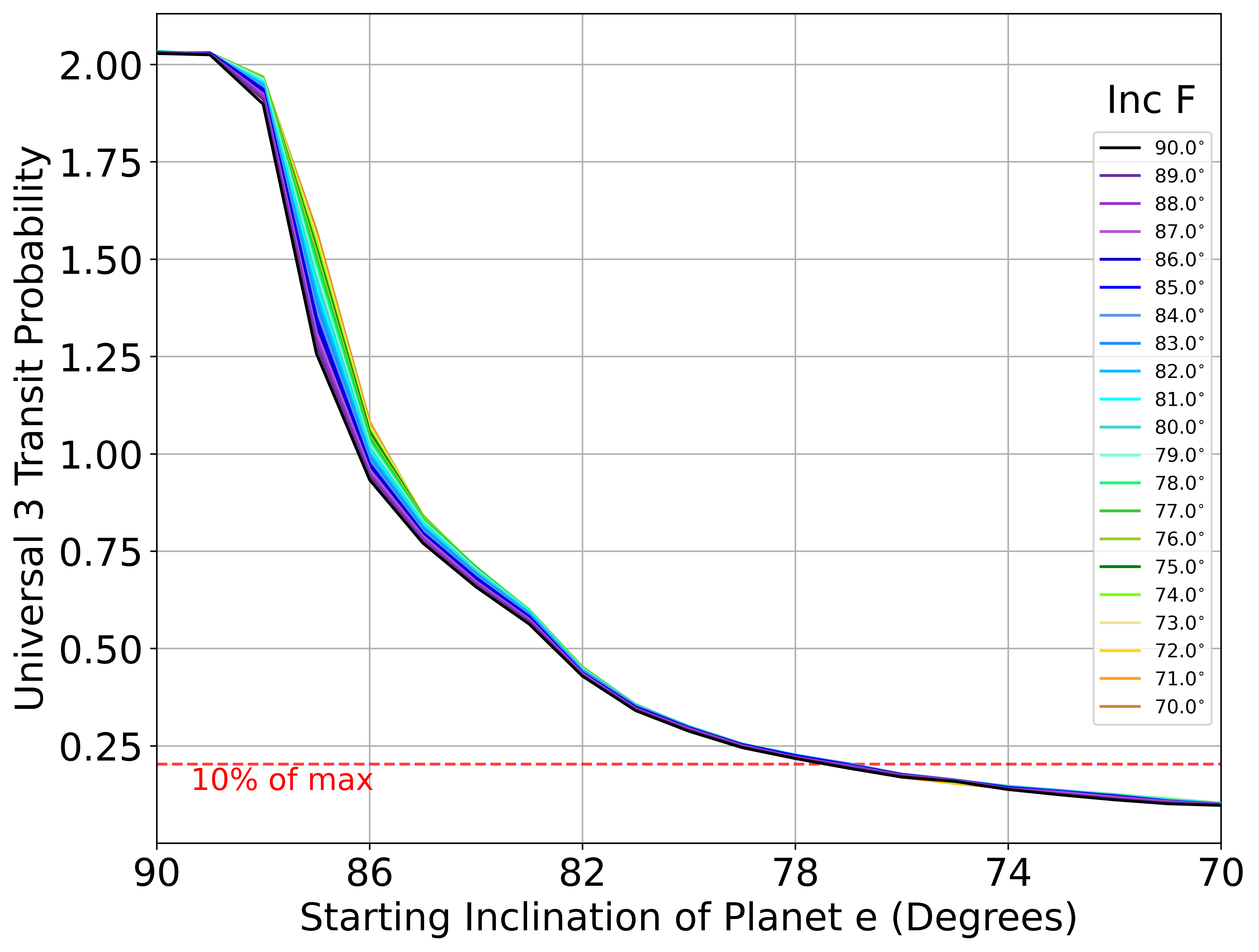}
    \caption{The universal probability of a randomly oriented observer seeing all three inner planets as transiting. As planet e is removed from the shared plane of the inner three, the probability decreases.}
    \label{fig:marginalizedTransitProb}
\end{figure}

 First interpreting Figure \ref{fig:marginalizedTransitProb}, we see that in the simulations in which planets e and f are both initialized at i = 90$^{\circ}$ inclination with respect to the line of sight, the 3-transit probability is equal to  the transit probability planet d alone (2.05\% as calculated by $p \sim R_* / a$), the outermost of the transiting three. Then, as planet e tilted away from the plane of the inner three planets, the universal three transit probability falls off, down to below 10\% of its original value at $i_e = 78^{\circ}$. This represents our constraint on the maximum mutual inclination, \coplanar, for planet e with respect to the shared plane of the three transiting planets.

\par Second, the aligned measurement of planet b's spin-orbit angle, along with the coplanar configuration derived from the mutual transit probability test above, implies that planets c, d, and e are also similarly well-aligned to the host star.

\par Third, through our dynamical simulations, we present the hypothesis that, while planet f does not play a role in the mutual transiting nature of the inner three planets, it is nevertheless most likely to be also in nearly the same shared plane as the inner four planets. The evidence for this argument, again, comes from Laplace-Lagrange secular perturbation theory \citep{LaplaceOriginal}. Planet f, being massive and very distant from the inner system, primarily torques the inner system as a whole, which then moves together as if it were a single rigid disk. This makes it so that the universal three transit probability is nearly the same for all starting angles of planet f. For a given value of planet e's initial inclination, there is a spread in probabilities with lower values of planet f providing marginally larger probabilities than higher values. These spreads in probabilities are at most $\sim0.05$\% and are therefore negligible. As planet f is inclined, the inner system precesses so that while it becomes unlikely for any one observer to see all three inner planets as transiting, the probability that at any given time, somewhere, someone sees all three as transiting is essentially constant, given a value of planet e's starting inclination. With this in mind, the precessing of the inner system would most likely express itself to us as a misaligned system. Given results above that the inner system is aligned, it is therefore most likely that the system is experiencing little precession from the influence of planet f, a similar argument was made in \citet{Kaib2011} for the 55 Cnc system.

\par If planet f was highly inclined and the inner system was precessing over the 85,000 year eigenfrequency, it would be unlikely that we would catch the system at a point where the observed RM effect points to an aligned orbit of planet b. Broadly speaking, planet f's inclination sets the average inclination of the plane roughly shared by the inner system. Then, the inner system's plane oscillates between that average value and $\pm$ the $\delta$ inclination of planet f with respect to 90$^{\circ}$. Therefore, to within an order of magnitude calculation, the probability of the inner system being aligned to within 10$^{\circ}$ (the 1$\sigma$ uncertainty on our spin-orbit measurement), given an inclination value of planet f, is equal to 10/(90-$i_f$). For the largest inclination for planet f tested in this work, 70$^{\circ}$, this probability is only 50\%. Therefore, the aligned RM measurement also loosely constrains planet f's inclination to be closer to 90$^{\circ}$ than not. Ultimately, this creates a picture of the HD 191939 system as both entirely flat and well-aligned, similar to our own solar system. Given the star's distance of 54 pc, it's mass, as well as planet f's mass and orbital separation, the astrometric signal of planet f is expected to be $\sim200$ micro-arcseconds. This is within the detection capabilities of Gaia \citep{Gaia2016} and future releases may detect the signal and give us greater insights into the orientation of the planet and therefore the system as a whole.

\par Our measurement of the obliquity of the multi-planet HD 191939 adds to a small list of only now 33 planets in multis that have had RM measurements performed. \citep{Southworth2011, Albrecht2022}. Even fewer of these have had the coplanarity of their system more fully explored through RM measurements of multiple planets in the same system. Of the five such measurements, three find all planets aligned: \citet{Hirano2020} found planets b, e, and f the TRAPPIST-1 system to all be aligned, TOI-942 b \citep{Wirth2021} and c \citep{Teng2024} are both aligned, and V1298 Tau b \citep{Marshall2022, Gaidos2022} and c \citep{Feinstein2021} are both well-aligned. Meanwhile \citet{Hjorth2021} finds two planets, one sub-Neptune and the other a Jovian, to be both misaligned to the star but aligned to each other and lastly \citet{Bourrier2021} finds planets b and c of the HD 3167 system to be on perpendicular orbits. Either HD 191939 c or d would be an excellent target to pursue an additional RM measurement in an effort to both add to this small list and confirm our dynamical results. Both planets are similar in radius to planet b and have similar impact parameters so their RM amplitudes are comparable and therefore attainable with KPF.

\par More planets in multi-planet systems have been had their obliquities measured through less precise methods than the RM effect, finding a variety of alignment and misalignment. These methods, such as spot-crossing \citep{Chaplin2013} and astroseismology \citep{Huber2013}, rely strictly on photometry and therefore are more amenable to measuring obliquities en masse. \citet{Dai2018} measured 60+ obliquities, mostly of single Hot Jupiter hosts using a spot-crossing technique and \citet{Campante2016} measured 25 obliquities by determining stellar inclination via astroseismology. Similarly, \citet{Louden2021} used ensemble measurements of $v\sin i$ to determine high and low obliquity systems with respect to the sky-plane, finding for hot stars ($>6250$K) the obliquity distribution of small planets is consistent with random stellar inclinations, or high obliquities. A follow up to this study, \citet{Louden2024}, with a larger sample size from TESS discoveries confirmed the earlier result and more solidly describes high system obliquity as a function of stellar temperature rather than of planet size.

\par Our RM measurement of HD 191939 b as well as the dynamical arguments for the broader alignment of the system bring a new and important data point to the census of obliquities of multi-planet systems. Our methods bridge the gap between precise obliquities from RM measurements and system-wide obliquities from other techniques. Through the three points above, the present day architecture of the HD 191939 system points to a formation scenario where all of the planets, including the distant super-Jovian, form out of the same aligned disk and each born with low mutual inclinations and eccentricities. In all, this leads to a relatively quiet dynamical history consistent with the Nebular Theory of planet formation \citep{Kant1755}.

\section{Conclusion}
\label{Conclusion}

\par We measure the sky-projected spin-orbit angle of HD 191939 b, a sub-Neptune in a multi-planet system, to be \obliquity. While small planets in multi-planet systems are known to be ubiquitous, our measurement adds an important data point to the still small, but growing list of such planets with measured spin-orbit angles. This understudied metric for an over-represented population in our current exoplanet census is vital to understanding how the typical multi-planet forms and evolves. With ever more precise spectrographs on large mirror telescopes, we now have the ability to make such measurements for more targets within this population.

\par Additionally, with an extended baseline of observations, we constrain the mass and orbit of planet f to be $M\sin i$ of 2.88 $\pm$ 0.26 $M_J$ and 2898 $\pm$ 152 days. With these parameters, we included planet f in a suite of dynamical simulations which robustly constrain the inclination of the non-transiting planet e to be within \coplanar of the shared plane of the inner three sub-Neptunes.

\par Furthermore, through the aligned measurement of planet b and the dynamical behavior of the system, we extrapolate that all the planets, including the distant planets e and f, are likely similarly well-aligned to the host star. This study represents one of the first to describe a system that begins to resemble our own solar system in terms of architecture: small planets interior to large planets, all in roughly the same orbital plane and all well-aligned to the host star. Extending this kind of detailed analysis to more systems will begin to allow us to probe how common or rare our solar system is in context of the known exoplanet census, eventually leading to a measurement of $\eta_{SS}$, the occurrence rate of solar systems.

\section{Acknowledgements}
\label{Acknowledgements}

\par We recognize and acknowledge the cultural role and reverence that the summit of Maunakea has within the indigenous Hawaiian community. We are deeply grateful to have the opportunity to conduct observations from this mountain.
\par J.L. and E.P recognize support from the Heising-Simons Foundation, grant number 2022-3833. This research has made use of the NASA Exoplanet Archive, which is operated by the California Institute of Technology, under contract with the National Aeronautics and Space Administration under the Exoplanet Exploration Program. This research made use of the open source Python package exoctk, the Exoplanet Characterization Toolkit \citep{exoctk}.
\par This research was carried out, in part, at the Jet Propulsion Laboratory and the California Institute of Technology under a contract with the National Aeronautics and Space Administration and funded through the President’s and Director’s Research \& Development Fund Program.

\begin{figure*}[t]
\centering\includegraphics[width=0.75\textwidth]{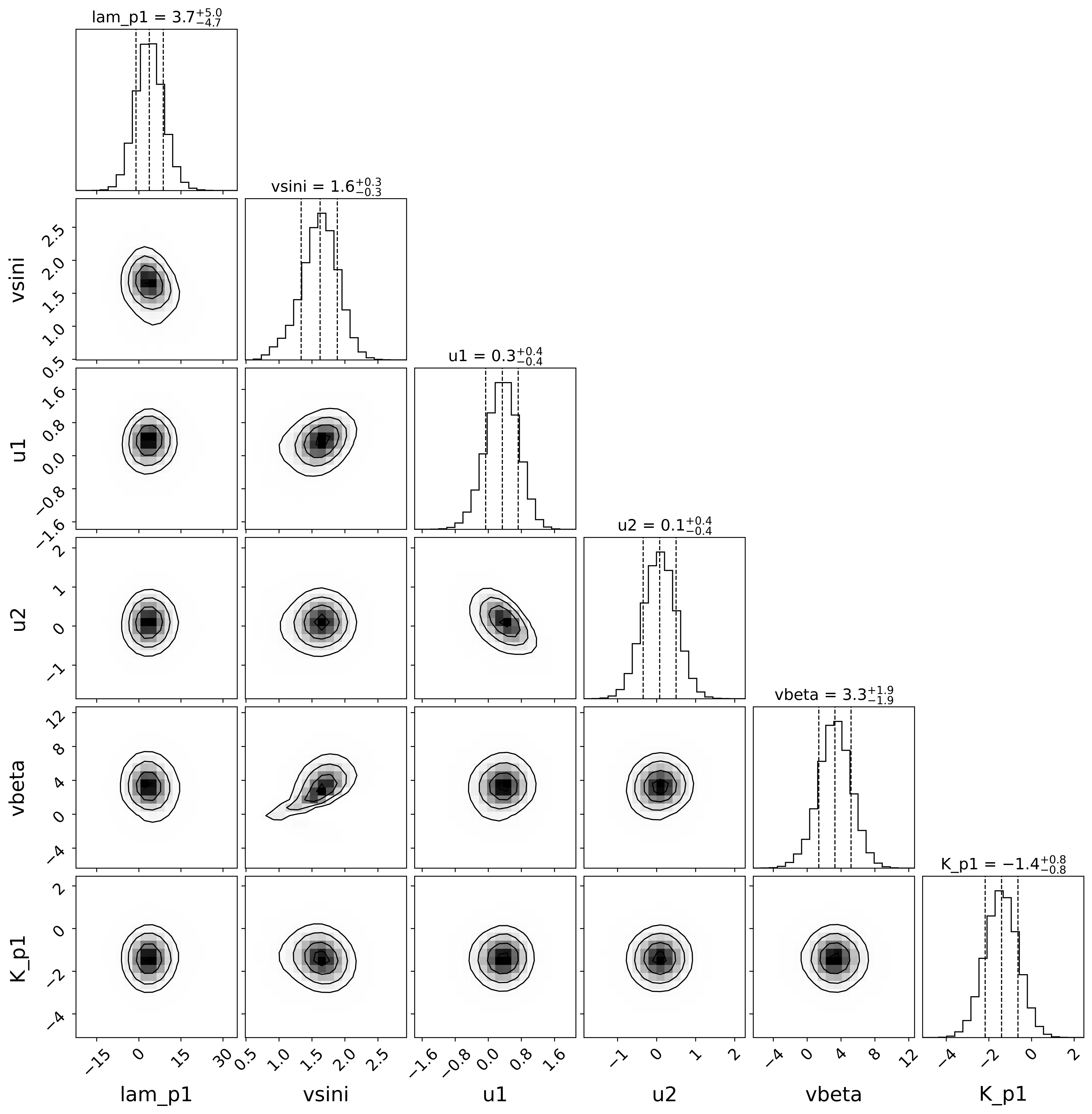}
    \caption{A corner plot of the posterior from MCMC sampling of the RM model.}
    \label{fig:RM_corner}
\end{figure*}

\bibliography{bib.bib}

\end{document}